\documentclass[runningheads]{llncs}
\usepackage{cite}
\usepackage{amsmath,amssymb,amsfonts}
\usepackage{graphicx}
\usepackage{textcomp}
\usepackage{xcolor}
\usepackage{listings, lstautogobble}
\usepackage{listings}
\usepackage{algorithm}
\usepackage{algpseudocode}
\usepackage{ragged2e}
\usepackage{array}
\usepackage{ifpdf}
\graphicspath{ {./images/} }
\def\BibTeX{{\rm B\kern-.05em{\sc i\kern-.025em b}\kern-.08em
    T\kern-.1667em\lower.7ex\hbox{E}\kern-.125emX}}

\newcommand{\Sysname}{PAVER}
\usepackage[T1]{fontenc}
\usepackage{graphicx}
\begin{document}
\title{Path-wise Vulnerability Mitigation}
%
%
\author{Zhen Huang\inst{1} \and Hiristina Dokic\inst{2}}

\institute{DePaul University, Chicago IL, USA \\\email{zhen.huang@depaul.edu} \and
DePaul University, Chicago IL, USA \\\email{hdjokic@depaul.edu}}
%
\maketitle              
\begin{abstract}
Software vulnerabilities are prevalent but fixing software vulnerabilities is not trivial. Studies have shown that a considerable pre-patch window exists because it often takes weeks or months for software vendors to fix a vulnerability. Existing approaches aim to reduce the pre-patch window by generating and applying mitigation patches that prevent adversaries from exploiting vulnerabilities rather than fix vulnerabilities. Because mitigation patches typically terminate the execution of vulnerability-triggering program paths at the level of functions, they can have significant side-effects. This paper describes an approach called \Sysname{} that generates and inserts mitigation patches at the level of program paths, i.e. path-wise vulnerability mitigation patches, in order to reduce their side-effects. \Sysname{} generates a  program path graph that includes the paths leading to vulnerabilities and the control dependencies on these paths, then identifies candidate patch locations based on the program path graph. For each candidate patch location, \Sysname{} generates and inserts a mitigation patch, and tests the patched program to assess the side-effect of the patch. It ranks the patches by the extent of their side-effects. We evaluates the prototype of \Sysname{} on real world vulnerabilities and the evaluation shows that our path-wise vulnerability mitigation patches can achieve minimum side-effects.

\end{abstract}
\newcommand{\nmappings}{11314}
\newcommand{\napps}{1260}


\renewcommand{\dbltopfraction}{1.0}
\renewcommand{\topfraction}{1.0}
\renewcommand{\bottomfraction}{1.0}
\renewcommand{\textfraction}{0.2}



\newcommand{\myfig}[5]
{
\begin{figure}[t]
\begin{center}
\ifpdf
\includegraphics[width=#4\linewidth]{#1}
\else
\includegraphics[width=#4\linewidth]{#1}
\fi
\end{center}
\vspace{#5}
\caption{#2}\label{#3}
\end{figure}
}

\newcommand{\myfigwide}[5]
{
\begin{figure*}[t]
\begin{center}
\ifpdf
\includegraphics[width=#4\linewidth]{#1}
\else
\includegraphics[width=#4\linewidth]{#1}
\fi
\end{center}
\vspace{#5}
\caption{#2}\label{#3}
\end{figure*}
}

\newcommand{\mysubfigtwo}[8]
{
\begin{figure}
        \centering
        \begin{subfigure}[b]{0.25\textwidth}
                \centering
                \includegraphics[width=\textwidth]{#1}
                \caption{#2}
                \label{#3}
        \end{subfigure}%
        ~ 
        \begin{subfigure}[b]{0.25\textwidth}
                \centering
                \includegraphics[width=\textwidth]{#4}
                \caption{#5}
                \label{#6}
        \end{subfigure}
        ~ 
        \caption{#7}\label{#8}
        \vspace{-15pt}        
\end{figure}
}

\newcommand{\myfigthreevert}[9]
{
\def\tempa{#1}
\def\tempb{#2}
\def\tempc{#3}
\def\tempd{#4}
\def\tempe{#5}
\def\tempf{#6}
\def\tempg{#7}
\def\temph{#8}
\def\tempi{#9}
\myfigthreevertcont
}

\newcommand{\myfigthreevertcont}[2]
{
\begin{figure}
        \centering
         \begin{tabular}{c}
        \begin{subfigure}[b]{0.6\textwidth}
                \centering
                \includegraphics[width=\textwidth]{\tempa}
                \caption{\tempb}
                \label{\tempc}
        \end{subfigure}%
	\\
	\\
	~
        \begin{subfigure}[b]{0.6\textwidth}
                \centering
                \includegraphics[width=\textwidth]{\tempd}
                \caption{\tempe}
                \label{\tempf}
        \end{subfigure}%
	\\
	\\
	~
        \begin{subfigure}[b]{0.6\textwidth}
                \centering
                \includegraphics[width=\textwidth]{\tempg}
                \caption{\temph}
                \label{\tempi}
        \end{subfigure}%
	\\
	\\
	\end{tabular}
	~
	\caption{#1}\label{#2}
        \vspace{-5pt}        
\end{figure}
}
\newcommand{\myfigthreevertbox}[9]
{
\def\tempa{#1}
\def\tempb{#2}
\def\tempc{#3}
\def\tempd{#4}
\def\tempe{#5}
\def\tempf{#6}
\def\tempg{#7}
\def\temph{#8}
\def\tempi{#9}
\myfigthreevertboxcont
}

\newcommand{\myfigthreevertboxcont}[2]
{
\begin{figure}
        \centering
         \begin{tabular}{c}
        \begin{subfigure}[b]{0.6\textwidth}
                \centering
                \setlength{\fboxsep}{0pt}%
                \setlength{\fboxrule}{1pt}%
                \fbox{\includegraphics[width=\textwidth]{\tempa}}%
                \caption{\tempb}
                \label{\tempc}
        \end{subfigure}%
	\\
	\\
	~
        \begin{subfigure}[b]{0.6\textwidth}
                \centering
                \setlength{\fboxsep}{0pt}%
                \setlength{\fboxrule}{1pt}%
                \fbox{\includegraphics[width=\textwidth]{\tempd}}%
                \caption{\tempe}
                \label{\tempf}
        \end{subfigure}%
	\\
	\\
	~
        \begin{subfigure}[b]{0.6\textwidth}
                \centering
                \setlength{\fboxsep}{0pt}%
                \setlength{\fboxrule}{1pt}%
                \fbox{\includegraphics[width=\textwidth]{\tempg}}%
                \caption{\temph}
                \label{\tempi}
        \end{subfigure}%
	\\
	\\
	\end{tabular}
	~
	\caption{#1}\label{#2}
        \vspace{-5pt}        
\end{figure}
}

\newcommand{\myfigtwovert}[8]
{
\begin{figure}
        \centering
         \begin{tabular}{c}
        \begin{subfigure}[b]{0.6\textwidth}
                \centering
                \includegraphics[width=\textwidth]{#1}
                \caption{#2}
                \label{#3}
        \end{subfigure}%
        \\
	\\
        ~ 
        \begin{subfigure}[b]{0.6\textwidth}
                \centering
                \includegraphics[width=\textwidth]{#4}
                \caption{#5}
                \label{#6}
        \end{subfigure}
        \end{tabular}
        ~ 
        \caption{#7}\label{#8}
        \vspace{-10pt}        
\end{figure}
}

\newcommand{\mysubfigtwovert}[8]
{
\begin{figure}
         \begin{tabular}{c}
        \begin{subfigure}[b]{0.3\textwidth}
                \centering
                \includegraphics[width=\textwidth]{#1}
                \caption{#2}
                \label{#3}
        \end{subfigure}%
        \\
        ~ 
        \begin{subfigure}[b]{0.3\textwidth}
                \centering
                \includegraphics[width=\textwidth]{#4}
                \caption{#5}
                \label{#6}
        \end{subfigure}
        \end{tabular}
        ~ 
        \caption{#7}\label{#8}
        \vspace{-10pt}        
\end{figure}
}

\newcommand{\mysubfigthree}[9]
{
\def\tempa{#1}
\def\tempb{#2}
\def\tempc{#3}
\def\tempd{#4}
\def\tempe{#5}
\def\tempf{#6}
\def\tempg{#7}
\def\temph{#8}
\def\tempi{#9}
\mysubfigthreecont
}

\newcommand{\mysubfigthreecont}[2]
{
\begin{figure*}
        \centering
        \begin{subfigure}[b]{0.3\textwidth}
                \centering
                \includegraphics[width=\textwidth]{\tempa}
                \caption{\tempb}
                \label{\tempc}
        \end{subfigure}%
        ~ 
        \begin{subfigure}[b]{0.3\textwidth}
                \centering
                \includegraphics[width=\textwidth]{\tempd}
                \caption{\tempe}
                \label{\tempf}
        \end{subfigure}
        \begin{subfigure}[b]{0.3\textwidth}
                \centering
                \includegraphics[width=\textwidth]{\tempg}
                \caption{\temph}
                \label{\tempi}
        \end{subfigure}
        ~ 
        \caption{#1}\label{#2}
        \vspace{-5pt}
\end{figure*}
}

\newcommand{\mysubfigthreebox}[9]
{
\def\tempa{#1}
\def\tempb{#2}
\def\tempc{#3}
\def\tempd{#4}
\def\tempe{#5}
\def\tempf{#6}
\def\tempg{#7}
\def\temph{#8}
\def\tempi{#9}
\mysubfigthreeboxcont
}

\newcommand{\mysubfigthreeboxcont}[2]
{
\begin{figure*}
        \centering
        \begin{tabular}{ccc}
        \begin{subfigure}[b]{0.32\textwidth}
                \centering
                {%
                \setlength{\fboxsep}{0pt}%
                \setlength{\fboxrule}{1pt}%
                \fbox{\includegraphics[width=\textwidth]{\tempa}}%
                }%
                \caption{\tempb}
                \label{\tempc}
        \end{subfigure}%
        &
        \begin{subfigure}[b]{0.32\textwidth}
                \centering
                {%
                \setlength{\fboxsep}{0pt}%
                \setlength{\fboxrule}{1pt}%
                \fbox{\includegraphics[width=\textwidth]{\tempd}}%
                }%
                \caption{\tempe}
                \label{\tempf}
        \end{subfigure}
        &
        \begin{subfigure}[b]{0.32\textwidth}
                \centering
                {%
                \setlength{\fboxsep}{0pt}%
                \setlength{\fboxrule}{1pt}%
                \fbox{\includegraphics[width=\textwidth]{\tempg}}%
                }%
                \caption{\temph}
                \label{\tempi}
        \end{subfigure}
        \end{tabular}
        \caption{#1}\label{#2}
        \vspace{-10pt}
\end{figure*}
}

\newcommand{\mysubfigfourbox}[9]
{
\def\tempa{#1}
\def\tempb{#2}
\def\tempc{#3}
\def\tempd{#4}
\def\tempe{#5}
\def\tempf{#6}
\def\tempg{#7}
\def\temph{#8}
\def\tempi{#9}
\mysubfigfourboxcont
}

\newcommand{\mysubfigfourboxcont}[5]
{
\begin{figure*}
        \centering
        \begin{tabular}{cc}
        \begin{subfigure}[b]{0.5\textwidth}
                \centering
                {%
                \setlength{\fboxsep}{0pt}%
                \setlength{\fboxrule}{1pt}%
                \fbox{\includegraphics[width=\textwidth]{\tempa}}%
                }%
                \caption{\tempb}
                \label{\tempc}
        \end{subfigure}%
        &
        \begin{subfigure}[b]{0.5\textwidth}
                \centering
                {%
                \setlength{\fboxsep}{0pt}%
                \setlength{\fboxrule}{1pt}%
                \fbox{\includegraphics[width=\textwidth]{\tempd}}%
                }%
                \caption{\tempe}
                \label{\tempf}
        \end{subfigure}
        \\
        \begin{subfigure}[b]{0.5\textwidth}
                \centering
                {%
                \setlength{\fboxsep}{0pt}%
                \setlength{\fboxrule}{1pt}%
                \fbox{\includegraphics[width=\textwidth]{\tempg}}%
                }%
                \caption{\temph}
                \label{\tempi}
        \end{subfigure}
        &
        \begin{subfigure}[b]{0.5\textwidth}
                \centering
                {%
                \setlength{\fboxsep}{0pt}%
                \setlength{\fboxrule}{1pt}%
                \fbox{\includegraphics[width=\textwidth]{#1}}%
                }%
                \caption{#2}
                \label{#3}
        \end{subfigure}
        \end{tabular}
        \caption{#4}\label{#5}
        \vspace{-10pt}
\end{figure*}
}

\newcommand{\mysubfigsixbox}[9]
{
\def\tempa{#1}
\def\tempb{#2}
\def\tempc{#3}
\def\tempd{#4}
\def\tempe{#5}
\def\tempf{#6}
\def\tempg{#7}
\def\temph{#8}
\def\tempi{#9}
\mysubfigsixboxcont
}

\newcommand{\mysubfigsixboxcont}[9]
{
\def\tempj{#1}
\def\tempk{#2}
\def\templ{#3}
\def\tempm{#4}
\def\tempn{#5}
\def\tempo{#6}
\def\tempp{#7}
\def\tempq{#8}
\def\tempr{#9}
\mysubfigsixboxcontcont
}

\newcommand{\mysubfigsixboxcontcont}[2]
{
\begin{figure*}
        \centering
        \begin{tabular}{ccc}
        \begin{subfigure}[b]{0.32\textwidth}
                \centering
                {%
                \setlength{\fboxsep}{0pt}%
                \setlength{\fboxrule}{1pt}%
                \fbox{\includegraphics[width=\textwidth]{\tempa}}%
                }%
                \caption{\tempb}
                \label{\tempc}
        \end{subfigure}%
        &
        \begin{subfigure}[b]{0.32\textwidth}
                \centering
                {%
                \setlength{\fboxsep}{0pt}%
                \setlength{\fboxrule}{1pt}%
                \fbox{\includegraphics[width=\textwidth]{\tempd}}%
                }%
                \caption{\tempe}
                \label{\tempf}
        \end{subfigure}
        &
        \begin{subfigure}[b]{0.32\textwidth}
                \centering
                {%
                \setlength{\fboxsep}{0pt}%
                \setlength{\fboxrule}{1pt}%
                \fbox{\includegraphics[width=\textwidth]{\tempg}}%
                }%
                \caption{\temph}
                \label{\tempi}
        \end{subfigure}
        \\
        \begin{subfigure}[b]{0.32\textwidth}
                \centering
                {%
                \setlength{\fboxsep}{0pt}%
                \setlength{\fboxrule}{1pt}%
                \fbox{\includegraphics[width=\textwidth]{\tempj}}%
                }%
                \caption{\tempk}
                \label{\templ}
        \end{subfigure}
        &
        \begin{subfigure}[b]{0.32\textwidth}
                \centering
                {%
                \setlength{\fboxsep}{0pt}%
                \setlength{\fboxrule}{1pt}%
                \fbox{\includegraphics[width=\textwidth]{\tempm}}%
                }%
                \caption{\tempn}
                \label{\tempo}
        \end{subfigure}
        &
        \begin{subfigure}[b]{0.32\textwidth}
                \centering
                {%
                \setlength{\fboxsep}{0pt}%
                \setlength{\fboxrule}{1pt}%
                \fbox{\includegraphics[width=\textwidth]{\tempp}}%
                }%
                \caption{\tempq}
                \label{\tempr}
        \end{subfigure}
        \end{tabular}
        \vspace{-8pt}
        \caption{#1}\label{#2}
\end{figure*}
}

\newcommand{\BA}{{\em begin\_atomic}}
\newcommand{\EA}{{\em end\_atomic}}

\newcounter{claimcounter}[section]
\newcolumntype{R}[1]{>{\raggedleft\arraybackslash}m{#1}}
\newcommand{\scbf}[1]{\vspace {0.05in}\noindent{\textbf{#1.}}}

\newcommand{\algorithmicbreak}{\textbf{break}}
\newcommand{\Break}{\State \algorithmicbreak}
\newcommand{\algorithmiccontinue}{\textbf{continue}}
\newcommand{\Continue}{\State \algorithmiccontinue}

\algrenewcomment[1]{\(\triangleright\) #1}
\algnewcommand{\LineComment}[1]{\State \(\triangleright\) #1}

\renewcommand{\algorithmicrequire}{\textbf{Input:}}
\renewcommand{\algorithmicensure}{\textbf{Output:}}

\section{Introduction}
Software vulnerabilities are weaknesses in computer systems that can be exploited by adversaries to mount cyberattacks, such as gaining unauthorized access to computer systems and stealing sensitive information. Due to the continuous demand for new functionality, software developers often prioritize adding new functionality over ensuring code security. This has led to an increasing number of vulnerabilities over the years, despite decades of effort in detecting vulnerabilities~\cite{cowan1998stackguard,ruwase2004practical,wang2010taintscope,li2016vulpecker,wu2017vulnerability,li2018vuldeepecker,lin2020software,xiao2020,HuangY21,chakraborty2021deep,AumpansubH21,li2021vuldetint,AumpansubH22,hin2022linevd,li2022path,CSR2022}. Over 26,000 software vulnerabilities were publicly reported in 2022~\cite{cvedetails} and the Common Vulnerabilities and Exposures (CVE) list currently contains more than 176,000 entries~\cite{vul}.


Particularly, many vulnerabilities published in recent years are severe because they can be exploited remotely. The percentage of remotely exploitable vulnerabilities have gradually increased to over 80\% of all the vulnerabilities~\cite{remote}. When a severe vulnerability is discovered, it is urgent to fix it. Unfortunately, it is not trivial to fix software vulnerabilities. It often takes weeks or months for software vendors to create and release patches for vulnerabilities because fixing a vulnerability typically requires multiple code changes~\cite{Talos,Iannone2023}. This creates a pre-patch window and gives adversaries plenty of opportunities to exploit known but yet patched vulnerabilities.

To reduce the pre-patch window, techniques have been proposed to mitigate vulnerabilities~\cite{Talos,RVM,undo}, in which they generate mitigation patches to prevent vulnerabilities from being exploited rather than to fix vulnerabilities. Because these kinds of patches serve a different purpose, they can be designed to be relatively simple and can be automatically generated. This allows them to be released quickly so that software users can apply them to defend against vulnerability exploits.

However, existing mitigation patches either have side-effects or must be customized for target software. Talos~\cite{Talos} and its successor RVM~\cite{RVM} generate mitigation patches to prevent vulnerable functions from being executed. Their mitigation patches are in the form of a return statement inserted at the beginning of vulnerable functions. Such a patch stops the execution of an entire vulnerable function and thus is effective in stopping exploits from triggering vulnerabilities, but it also causes programs to lose all the functionality provided by the vulnerable function. It stops the execution of all the paths going through the vulnerable paths, regardless of whether these paths can lead to a vulnerability. If the vulnerable function happens to be on the critical path of a program, the program is essentially rendered unusable. 

This paper presents an approach that generates mitigation patches with minimized side-effects. Our insight is that a mitigation patch should stop vulnerable paths instead of vulnerable functions. By inserting a mitigation patch on only the program paths leading to a vulnerability, it stops the execution of these vulnerable program paths but still allows the execution of other program paths irrelevant to the vulnerability even when they involve the vulnerable functions. This way it can preserve the functionality irrelevant to the vulnerability. We call this approach \Sysname{}, short for PAth-wise VulnERability mitigation. 

\Sysname{} generates mitigation patches in three phases: finding vulnerable paths, identifying patch locations, and synthesizing and ranking patches. Phase 1 uses the call graph and control dependency graph of a target program to generate a program path graph. Phase 2 identifies patch locations on each path in the program path graph. Phase 3 synthesizes and inserts patches at each candidate patch location and then assesses the side-effect of the patches by testing the patched program. It ranks the patches by their side-effects.

In phase 1, it uses static analysis to find the paths leading to a vulnerability. For a target program, the static analysis works in two steps. First, it identifies all possible paths that can call the vulnerable function, based on the call graph of the target program. Second, it identifies the control dependencies that govern the execution of the paths. The two steps generate a program path graph containing all the paths from the program's entry point to the vulnerable statement, as well as information on control dependencies.


\myfig{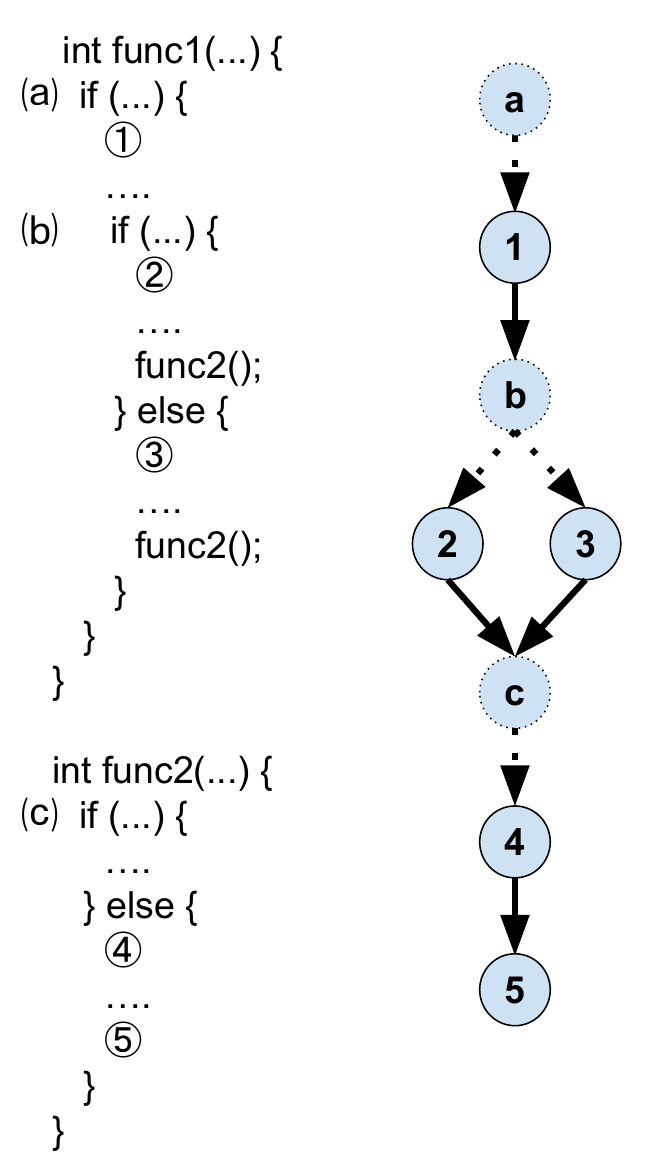}{Example path-wise vulnerability mitigation patch locations: both path a-1-b-2-c-4-5 and path a-1-b-3-c-4-5 lead to a vulnerability; basic block 1, 4, or basic blocks 2 and 3 are candidate patch locations; dotted arrows denote control dependencies.}{fig:patch_locations}{0.6}{0pt}


Phase 2 identifies all candidate patch locations. Given a path in the program path graph, it deems the first successor basic block of each conditional basic block as a candidate patch location because a conditional basic block governs the execution of all the basic blocks after itself on the path. For the example code in Figure~\ref{fig:patch_locations}, it will identify basic block 1, 4, or basic blocks 2 and 3 as candidate patch locations, as they are the first successors of conditional basic block a, c, and b respectively.

Essentially a mitigation patch inserted at each candidate patch location prevents an exploit to trigger a specific vulnerability. To reduce side-effects, our mitigation patch diverts program execution to error handling code rather than to terminate program execution. To synthesize a patch at a candidate patch location, phase 3 finds error handling code accessible at the location and creates a patch accordingly. As patches inserted at different patch locations can have different side-effects, phase 3 assess the side-effect of the patches, It applies each patch to the target program and tests the patched program. The test result indicates the extent of the side-effect of a patch, which is used as the basis to rank the patches.

This paper makes the following main contributions:
\begin{itemize}
    \item We propose an approach called \Sysname{} to mitigating vulnerabilities with minimum side-effects by inserting mitigation patches at the program paths leading to the vulnerabilities.
    \item We develop a technique to generate program path graphs and identify candidate patch locations.
    \item We have implemented the approach in a prototype and evaluated it on real world vulnerabilities. We describe our design and evaluation in the paper.
    \item Our evaluation shows that \Sysname{} can generate mitigation patches with minimum side-effects for real world vulnerabilities.
\end{itemize}

Our paper is structured as follows. Section~\ref{sec:related} discusses related work. Section~\ref{sec:problem} defines the problem addressed by this work. We present our design in Section~\ref{sec:design} and evaluation in Section~\ref{sec:evaluation}. We discuss the limitations of our work in Section~\ref{sec:discussion} and conclude in Section~\ref{sec:conclusion}.
\section{Problem Definition}\label{sec:problem}
In this work, we focus on vulnerability mitigation, a way to generate mitigation patches to address vulnerabilities rapidly. Because mitigation patches do not aim to fix vulnerabilities, they can have side-effects. The goal of this work is to design and generate mitigation patches that have minimum side-effects. 

\scbf{Vulnerability mitigation} We define vulnerability mitigation as a way to patch a program in order to prevent adversaries from exploiting vulnerabilities. This kind of patches are called \textit{mitigation patches}. Unlike regular patches aiming to fix vulnerabilities, mitigation patches are \textit{not} designed to fix vulnerabilities. They are intended as a rapid and temporary means to address vulnerabilities before the fix for vulnerabilities are available.




\scbf{Mitigation patch} A mitigation patch is a software update that changes program code to addresses software vulnerabilities. It protects programs from malicious attacks attempting to trigger vulnerabilities. Mitigation patches are designed to be simple and can be automatically generated. This design allows mitigation patches to be rapidly generated and released so that vulnerabilities can be addresses quickly. 

\scbf{Side-effect} A side-effect is an unintended consequence caused by a mitigation patch. Because a mitigation patch intends to prevent adversaries from exploiting a vulnerability, any impact on the behaviors of the target program that is irrelevant to the vulnerability can be considered as a side-effect. We define the side-effects of a mitigation patch as the extent of the functionality loss caused by the mitigation patch.

\section{Design}\label{sec:design}
We design \Sysname{} to generate patches for mitigating vulnerabilities. These mitigation patches disable the execution of vulnerable code to prevent adversaries from exploiting vulnerabilities. Because they disable code execution, they can have the side-effects of causing target programs to lose the functionality provided by the disabled code. Our goal is to minimize the side-effects of the patches. In other words, we aim to generate mitigation patches that can preserve the functionality of target programs. 

\subsection{Overview}
\Sysname{} produces mitigation patches for a vulnerability in three phases: 1) finding vulnerable paths, 2) identifying patch locations, and 3) synthesizing and ranking patches. Figure~\ref{fig:paver} illustrates the workflow of \Sysname{}.

\myfig{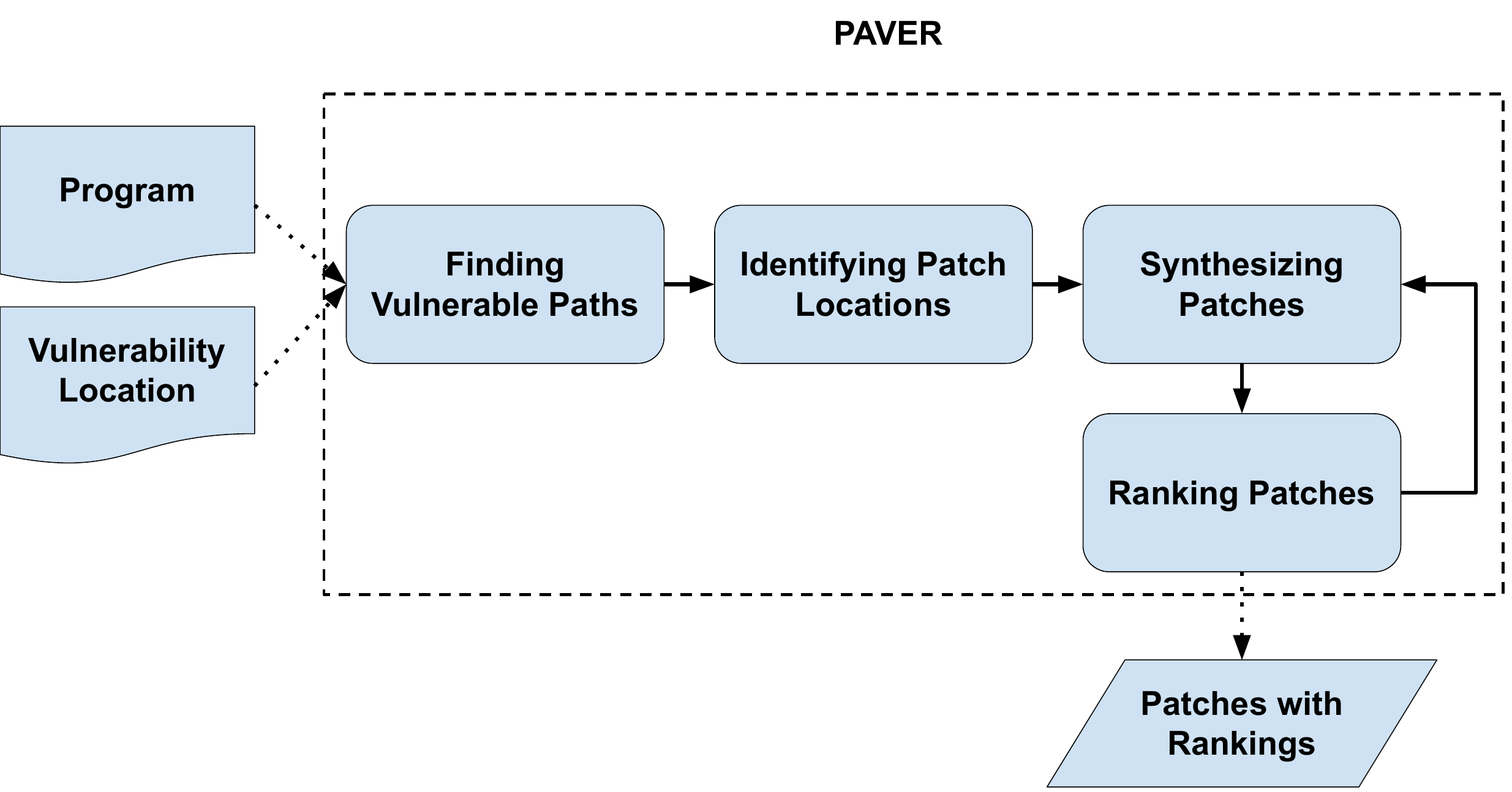}{Workflow of \Sysname{}.}{fig:paver}{1.0}{0pt}

\begin{enumerate}
\item \scbf{Finding vulnerable paths} Our mitigation patches mitigate vulnerabilities by terminating program paths leading to  vulnerabilities. To generate and insert a patch for a vulnerability, \Sysname{} first finds program paths reaching the vulnerability and the control dependencies that determine whether to execute the paths. This step outputs a program path graph that contains the paths and control dependencies.

\item \scbf{Identifying patch locations} Based on the program path graph, this step identifies locations on the paths that are suitable for inserting mitigation patches. For each conditional basic block on a path, we choose the first successor basic block of the conditional basic block as a patch location. As this successor basic block dominates the rest of the basic blocks on the path, inserting a patch at this location terminates the path as soon as the condition basic block causes the execution to take the path. This step outputs a list of candidate patch locations.

\item \scbf{Synthesizing and ranking patches} This step generates patches,  measures the side-effects of the patches, and ranks the patches by their side-effects. For each candidate patch location, it finds the error handling code that is accessible at the location and creates a patch leveraging the error handling code. It then applies the patch to the target program and runs the patched program against a set of test cases to measure the side-effects of the patch. Finally it ranks the patches based on the side-effects of each patch.

\end{enumerate} 

To illustrate how \Sysname{} generates patches, we use the vulnerability example in Listing~\ref{lst:list1} as the target vulnerability. The vulnerability, CVE-2017-9171, was discovered in Autotrace, an image processing tool. It is an out-of-bound read vulnerability, occurring in the \texttt{ReadImage} function. We focus on the part of the program paths inside the function for this illustration. 

\begin{lstlisting}[float,floatplacement=H,caption={Example vulnerability, adopted from an out-of-bound read vulnerability CVE-2017-9171 in AutoTrace.},captionpos=b, label={lst:list1}]
unsigned char* ReadImage (....) {
   unsigned char *image;
   
   if (bpp <= 8){
      unsigned char *temp2, *temp3;
      unsigned char i;
      temp2 = temp = image;
      XMALLOC (image, w * h * 3 * 
      sizeof (unsigned char));
      temp3 = image;
      for (ypos=0; ypos<h; ypos++) {
         for (xpos=0; xpos<w; xpos++) {
            index = *temp2++;
            *temp3++ = cmap[i][0];
            if (!grey) {
               *temp3++ = cmap[i][1];
               *temp3++ = cmap[i][2];
            }
         }
      }
      free (temp);
   }
   free (buffer);
   return image;
}
\end{lstlisting}


Phase 1 of \Sysname{} takes the information on the vulnerability as input, and outputs a program path graph. The input information needs to specify the location of the vulnerability, which can be a vulnerable statement or a vulnerable function. For the example vulnerability, we assume the input information to phase 1 specifies that line 16 is the vulnerable statement, which triggers an out-of-bound read via pointer \texttt{temp}. 

Given this vulnerability location, phase 1 will identify the path from the entry of the function to the vulnerable statement as 4-5-11-12-13-15-16, in which 4, 11, 12, and 15 are conditional statements. It produces a program path graph containing the path with labels on the conditional statements.

Phase 2 takes the program path graph generated by phase 1, and outputs a list of candidate patch locations. It iterates through each conditional statement on the path and identifies the successor of the conditional statement as a candidate patch location. If a successor is also a condition statement, it does not consider this successor as a candidate patch location. As a result, it will identify line 5, 13, and 16 as candidate patch locations for the example vulnerability, and output them as a list. 

In phase 3, \Sysname{} takes the list of candidate patch locations produced by phase 2 as input, and outputs a list of patches and the ranking of the patches. It goes through each candidate patch location in the list and synthesizes a patch, which is a \texttt{return} statement that returns an error return value, for this candidate patch location. For function \texttt{ReadImage}, \Sysname{} will identify \texttt{NULL} as its error return value for synthesizing patches for the candidate patch locations within the function.

After synthesizing a patch for a candidate patch location, phase 3 applies the patch to the target program and runs the patched program against a set of test cases. It uses the ratio of the number of passed test cases over the number of all the test cases, called preserved functionality ratio (FPR), as the metric of the side-effects of the patch.

This process of synthesizing and testing a patch repeats for all the candidate patch locations. Phase 3 keeps track of the FPR for each patch and ranks the patches by their FPRs, in the order from the highest FPR to the lowest FPR. It outputs the list of generated patches and their associated rankings.

\subsection{Finding Vulnerable Paths}
For a given vulnerability location, the program paths triggering the vulnerability can be found with either static analysis or dynamic analysis. The static analysis is typically conservative and can find all program paths, but it may have false positives. While the dynamic analysis is more accurate, it is challenging for the dynamic analysis to cover all program paths. 

We choose to use static analysis in this phase to cover all possible paths leading to a vulnerability location, in order to ensure that the vulnerability is mitigated completely. For a target program, our static analysis first generates a call graph of the program, and then uses the call graph to find the call chains to the function containing the vulnerability location, i.e. the vulnerable function. 

Based on the call graph, it uses a backward reachability analysis to find call chains, starting from the vulnerable function and going through the direct and indirect callers of the vulnerable function until the entry function of the target program. To take into account function calls via function pointers, it considers all the functions to which function pointers are taken and whose function prototypes match the types of function pointers as possible callees via these function pointers.

It then generates a control dependency graph of the program and finds the control dependencies governing the calls on the chains. For each call on a call chain, it finds all the conditional statements on which the call is control dependent. For the vulnerable function, it also finds the control dependencies governing the statements on the execution paths to the vulnerable statement.

This phase generates a program path graph consisting of program paths and the control dependencies governing the execution of the paths. Each program path is composed of the basic blocks from the first basic block of the entry function to the basic block containing the vulnerable statement. Each basic block on a program path is associated with a label indicating whether the basic block is a conditional basic block.

\subsection{Identifying Patch Locations}
A mitigation patch can be inserted into any basic block on a program path in the program path graph to terminate the execution of the path. However, some basic blocks on such a program path can also be part of the program paths not necessarily leading to the target vulnerability. For example, any conditional basic block on a program path will be executed regardless whether the outcome of the conditional basic block determines to continue on the path leading to the vulnerability or not. 

In order to minimize the side-effects of mitigation patches, the candidate patch locations must satisfy two requirements. First, the basic blocks should be likely to lead to the vulnerability. This requirement ensures that patches inserted at these locations are less likely to be executed for inputs that will not trigger the vulnerability. These include non-conditional basic blocks that are on the program path to the vulnerability. Second, the path should be terminated as early as possible. This is because the more code on a path is executed the more likely it will involve operations that need to be undone when terminating the path. 

To satisfy the two requirements, we choose the first non-conditional successor basic block of each conditional basic block on a program path as a candidate patch location. A patch inserted at such a location essentially terminates a program path as soon as the outcome of the conditional basic block governing the execution of the location turns out to continue the execution on the program path. 

One might think that the candidate patch location closet to the vulnerable statement would have the least side-effects because this location has the highest probability, among other candidate patch locations, to trigger the vulnerability. However, this is not necessarily the case. This is because 1) executing the vulnerable statement does not necessarily trigger the vulnerability and 2) the side-effects of a patch are dependent on how many paths go through the patch. Our evaluation in Section~\ref{sec:evaluation} also confirms that.

This phase outputs a list of candidate patch locations for all the paths leading to the vulnerability. Each patch location refers to a basic block. 

\subsection{Synthesizing and Ranking Patches}
\Sysname{} synthesizes patches that prevent adversaries from exploiting vulnerabilities by diverting program execution to error handling code. Each patch terminates the execution of a path leading to a vulnerability and leverages the existing error handling code to let the target program resume program execution. By inserting patches at candidate patch locations identified by phase 2, the patches are likely to be executed only for malicious inputs aiming to exploit vulnerabilities instead of benign inputs.

Like the patches generated by Talos, our patches are in the form of a \texttt{return} statement. This phase consists of five steps. First, \Sysname{} creates an empty patch list and adds all candidate patch locations into a working list.

Second, it retrieves and removes a candidate patch location from the working list, and synthesizes a patch for the location. It uses Talos to find existing error handling code accessible to the location, particularly a return value used by the function containing the location to indicate an error condition. With the identified error return value, \Sysname{} synthesizes a patch that returns the error return value to the caller of the function.

Third, \Sysname{} applies the patch to the candidate patch location in the target program. The patch can be applied in different representations of the target program, such as binary code or source code. If it is applied to the source code, the target program needs to be re-compiled.

Fourth, \Sysname{} then runs the patched program using a set of test cases. It keeps track of the number of passed test cases and computes the PFR for the patch as the ratio of the number of passed test cases over the number of all test cases. After that, it adds the patch location, the patch and its PFR to the patch list.

Last, it goes back to the second step to work on a another candidate patch location if the working list is not empty. If the working list is empty, it ranks the patches in the patch list by their PFRs. The patches are ranking from the one with the highest PFR to the one with the lowest PFR. The ranked patch list is the output of this phase.

\section{Evaluation}\label{sec:evaluation}
In this section, we present the evaluation results of our prototype of \Sysname{}. We evaluate it on different types of real world vulnerabilities in a variety of programs. The evaluation focuses on assessing the side-effects of the patches generated by our approach. Particularly we compare that with the side-effects of the patches generated by Talos~\cite{Talos}.

\subsection{Measuring Side-effects}
The side-effects of patches can be measured in different ways. As software users typically concern about the functionality affected by patches, we choose to use test cases to measure the side-effects because test cases are usually designed to evaluate a program at the level of functionality.

For each synthesized patch for a program, \Sysname{} applies the patch to the program and runs the patched program against a set of test cases for the program. It uses preserved functionality ratio (FPR), the ratio of the number of passed test cases over the number of all test cases, as the metric of the side-effects of the patch. As a result, we consider a patch has less side-effect if it has higher PFR.

An acute reader will find that the soundness of this metric is dependent on the quality of the test cases. The more comprehensive the test cases are, the more accurate the metric is. Because mature and popular programs tend to have high quality test cases, we choose vulnerabilities for our evaluation from such kinds of programs. 


\subsection{Benchmarks}
We find real world vulnerabilities from popular online
vulnerability databases and bug databases, including CVE
- MITRE~\cite{cve}, NVD~\cite{nvd}, and BugZilla~\cite{bugzilla}. We were able to reproduce 14 vulnerabilities of four different types from the four programs listed in Table~\ref{tbl:programs}. All the programs are commonly used and have been actively developed and maintained for years. The sizes of these programs range from 19,264 to 431,063 lines of source code.

\begin{table}
    \Centering
    \caption{List of evaluated programs. Column ``\# Vulns.'' presents the number of vulnerabilities from each program; column ``Size'' shows the size of each program in terms of its number of source code lines.}
    \begin{tabular}{|l|c|l|r|}
    \hline
    \textbf{Program} & \textbf{\# Vulns.} & \textbf{Description} & \textbf{Size}\\
     \hline
                \RaggedRight{autotrace} &
                \RaggedRight{6} &
                \RaggedRight{image processing tool}  & 19,264\\
                \RaggedRight{tiff} &
                \RaggedRight{4} &
                \RaggedRight{image processing tool} & 25,604\\
                \RaggedRight{php} &
                \RaggedRight{2} &
                \RaggedRight{programming language interpreter} & 340,491\\
                \RaggedRight{python} &
                \RaggedRight{2} &
                programming language interpreter & 431,063\\
                
    \hline   
    \end{tabular}
    \label{tbl:programs}
\end{table}

Table~\ref{tbl:vulnerabilities} lists the vulnerabilities. Column ``Type'' shows the type of each vulnerability, with A for assertion, B for buffer overflow, I for integer overflow, and O for out-of-bound read. We have one vulnerability causing an assertion, six buffer overflow vulnerabilities, four integer overflow vulnerabilities, and three vulnerabilities causing out-of-bound read.

\begin{table}
\caption{List of evaluated vulnerabilities. }
\centering
\begin{tabular}{ |l|c|l|c|r|r| } 
 \hline
 \textbf{CVE\#} & \textbf{Type} & \textbf{Program} & \textbf{\# Tests}  & \textbf{\Sysname{}} & \textbf{ Talos}\\
 \hline
 2017-9171 & O & autotrace & 87 & 85 (98\%) & 0\\ 
 2017-9172 & B & autotrace & 87 & 85 (98\%) & 0\\ 
 2017-9173 & B & autotrace & 87 & 85 (98\%)  & 0\\ 
 2017-9174 & O & autotrace & 87 & 85 (98\%) & 0 \\ 
 2017-9186 & A & autotrace & 87 & 85 (98\%) & 0 \\  
 2017-9189 & O & autotrace & 87 & 0 & 0 \\ 
 2006-2025 & I & tiff & 68 & 0 & 0\\
 2009-2285 & B & tiff & 68 & 37 (54\%) & 37 (54\%)\\
 2015-8668 & B & tiff & 56 & 56 (100\%) & 56 (100\%)\\
 2016-10095 & B & tiff & 68 & 68 (100\%) & 0 \\
 2007-1383 & I & php & 591 &   589 (99\%) & 588 (99\%) \\
 2013-7226 & I & php & 591 &  86 (15\%) & 10 (2\%) \\ 
 2014-1912 & B & python & 94 &  86 (94\%) & 86 (91\%) \\
 2016-5636 & I & python & 94 &  74 (79\%) & 71 (76\%) \\
 \hline
\end{tabular}
\label{tbl:vulnerabilities}
\end{table}


\subsection{Side-effects of Patches}
For each vulnerability, we run \Sysname{} to generate patches to mitigate it. After generating a patch, \Sysname{} applies the patch to its corresponding program, then runs the patched program against a proof-of-concept exploit to verify the effectiveness of the patch, and against test cases to measure the side-effects of the patch. \Sysname{} ranks all the patches it generated for a vulnerability by the number of passed test cases of the patched program.

All the patches generated by \Sysname{} for a vulnerability effectively mitigate the vulnerability. \Sysname{} verified that the proof-of-concept exploit for each vulnerability can no longer trigger the vulnerability after applying each patch. 

Column ``\# Tests''  in Table~\ref{tbl:vulnerabilities} presents the numbers of test cases we used for measuring the side-effects of the patches mitigating the vulnerabilities. They range from 56 to 591, with a median of 77 and a mean of 224.

Column ``\Sysname{}'' shows the number of passed test cases after applying the highest ranked patch generated by \Sysname{}. The percentage besides each number is the PFR for the patch. As a comparison, column ``Talos'' shows the number of passed test cases after applying the patch generated by Talos~\cite{Talos}. 

As we can see, the highest ranked patches generated by \Sysname{} have a decent side-effect for the vast majority of these vulnerabilities. Except for two vulnerabilities (CVE-2017-9189 and CVE-2006-2025), the patches generated by \Sysname{} have a PFR from 15\% to 100\%, with a median of 98\% and a mean of 86\%. For these two vulnerabilities, the patches fail all the test cases.

In contrast, the patches generated by Talos fail all the test cases for the majority of the vulnerabilities. Excluding these eight vulnerabilities, the patches generated by Talos have a PFR from 2\% to 100\%, with a median of 84\% and a mean of 70\%.

The results show that the patches generated by our path-wise vulnerability mitigation can preserve considerably more functionality than those generated by function-level vulnerability mitigation, which is used by Talos. And these patches achieve the same effect in mitigating vulnerabilities. As vulnerability mitigation is a trade-off between functionality and security, the patches preserve more functionality are more preferable, given that they provide the same level of security.


\subsection{Levels of Patches}\label{sec:side_effect_detail}
\Sysname{} generates patches at all the candidate patch locations for each vulnerability. Table~\ref{tbl:patches} presents the details on the patches generated by \Sysname{}. The number of patches generated for each vulnerability varies from 13 to 47.

Column ``\# Levels'' shows the maximum number of functions on the call chain in all the paths leading to each vulnerability. Column ``Best Patch Level'' presents the level of the function in which the highest ranked patch is inserted. The level starts from the vulnerable function. In other words, level 0 is the vulnerable function while level 1 is the direct caller of the vulnerable function.

We can see that the highest ranked patches for eight vulnerabilities exist at level 1 and the highest ranked patch for one vulnerability exists at level 0. For the other five vulnerabilities, most of the highest ranked patches exist in the middle of the levels. except for one vulnerability. Our conjecture is that the side-effects of a patch tend to be more limited if it is inserted into a low level function, unless the function is commonly used by many functions, such as a library function. We plan to test the conjecture in our future work.

\begin{table}
\centering
\caption{Patches generated by \Sysname{}.}
\begin{tabular}{ |l|r|r|r| } 
 \hline
 \textbf{CVE\#} & \textbf{\# Patches} & \textbf{\# Levels} & \textbf{Best Patch Level}\\
 \hline
 2017-9171 & 28 & 4 & 1\\
 2017-9172 & 28 & 4 & 1\\ 
 2017-9173 & 28 & 4 & 1\\
 2017-9174 & 28 & 4 & 1\\ 
 2017-9186 & 33 & 4 & 1\\ 
 2017-9189 & 22 & 7 & 1\\
 2006-2025 & 43 & 10 & 1\\
 2009-2285 & 29 & 6 & 4\\
 2015-8668 & 47 & 3 & 1\\
 2016-10095 & 13 & 4 & 0\\
 2007-1383 & 25 & 8 & 5\\
 2013-7226 & 22 & 10 & 6\\ 
 2014-1912 & 45 & 12 & 11\\
 2016-5636 & 23 & 10 & 6\\
  \hline
\end{tabular}
\label{tbl:patches}
\end{table}

\section{Related Work}\label{sec:related}
Vulnerability mitigation techniques can be broadly categorized into code patching~\cite{costa2007bouncer,RVM,Talos,improving,adaptive,perkins2009automatically,qian2020,EuroSec21,undo,Huang2022,ahmad2022trimmer,Huang_Tan_Yu_2022,mansouri2023eliminating}, and rule-based mitigation~\cite{wahbe1993efficient,abadi2009control,wang2004shield,criswell2014kcofi,instaguard,wang2023pet}.



\scbf{Code patching} These techniques generate vulnerability mitigation patches for the code of target programs. Many of them synthesize patches that become in effect when the patched programs are re-executed, similar to regular patches. RVM~\cite{RVM} and Talos~\cite{Talos} generate patches that are called Security Workaround for Rapid Response (SWRR), which prevents a vulnerability from being triggered by disabling the execution of an entire vulnerable function. Because a function can contain both program paths that are vulnerability-relevant and program paths that are \textit{not}, an SWRR disables all of them and can lead to the side effect of disabling vulnerability-irrelevant functionality. Different from an SWRR, each patch generated by \Sysname{} disables the execution of vulnerability-relevant program paths. This approach will reduce or eliminate the risk of disabling vulnerability-irrelevant functionality.

Hecaton~\cite{undo} generates bowknots for mitigating kernel bugs and vulnerabilities. A bowknot stops the execution of a system call that is about to trigger a bug and reverses the effects of the operations that have already been performed by the system call, in order to leave the kernel in a consistent state. This approach learns how to reverse system call operations by identifying specific coding patterns of system calls. In contrast, \Sysname{} relies on existing error handling code to maintain the consistency of program state. 

Other techniques synthesize patches that can be applied to programs that are being executed, and thus become in effect as soon as they have been applied. KARMA~\cite{adaptive} translates a kernel patch written in a high-level memory-safe language into a binary patch. By restricting the operations that can be performed by such a patch and the program locations where the patch can be inserted, KARMA enables the deployment of the binary patches to a running kernel. While KARMA performs the translation offline, INSIDER~\cite{improving} uses just-in-time compilation to translate patches written in C into binary patches. It deploys such a patch into a running program by applying the patch to a second copy of the program code and diverting the program execution to the patched copy.

\scbf{Rule-based mitigation} These techniques do not involve patching the code of target programs. Instead, they employ rules that are enforced at runtime to prevent the exploitation of vulnerabilities. Some techniques restrict the control flow of a target program to only legitimate transfers intended by the program, i.e. maintaining Control Flow Integrity (CFI)~\cite{abadi2009control}. They focus on preventing vulnerability exploits to hijack program executions. The legitimate control flow transfers of a target program are typically identified by analyzing the code of the program. Some techniques rely on the support of CPUs or OSes. InstaGuard~\cite{instaguard} translates a patch written in a simple language into a sequence of debugging primitives supported by ARM CPUs, including breakpoints, watchpoints, and assertions. These debugging primitives can be applied to a running program to check for vulnerability conditions and prevent vulnerabilities from being triggered, without changing the code of the program. By contrast, PET~\cite{wang2023pet} translates vulnerability conditions reported by Linux kernel sanitizers such as Kernel
Address Sanitizer (KASAN) and Kernel Memory Sanitizer (KMSAN) into eBPF filters that can be applied to a running kernel to detect exploits to kernel vulnerabilities.

\section{Limitations}\label{sec:discussion}
The mitigation patches generated by \Sysname{} is simply a \texttt{return} statement that returns an error return value for the function containing the patch. Due to its simplicity, \Sysname{} can synthesize patches in a straightforward way. However, the simple patches do not work for the scenarios where the code before the patch has already executed some operations that must be undone before the function returns. For these scenarios, the patches may cause significant side-effects such as unexpected behaviors. We plan to address the limitation by finding the operations that need to be undone and adding the operations in patches to undo them, a technique used by bowknots~\cite{undo}.

Our prototype of \Sysname{} automatically synthesizes source code patches and applies them to the source code of a target program. As patch locations are often part of some conditional statements, it needs to find the first line in the correct branch of these conditional statements. It uses heuristics and simple source code parsing to find the correct source code locations to insert the patches. This approach works for the coding styles of the evaluated programs, but it may not work for other coding styles. One way to improve the approach is to synthesize patches in the form of LLVM bitcode and insert them into the LLVM bitcode of a target program.

\section{Conclusion}\label{sec:conclusion}
Vulnerability mitigation is a promising approach to rapidly preventing adversaries from exploiting known but yet patched vulnerabilities. This paper presents \Sysname{}, an approach that automatically generates and inserts mitigation patches at program paths leading to vulnerabilities. These mitigation patches let these program paths fall back to error handling code before triggering vulnerabilities, aiming to minimize the side-effects caused by changing the program paths. \Sysname{} first generates a program path graph that contains the program paths causing vulnerabilities and the control dependencies that govern the execution of the paths. Based on the program path graph, it then identifies candidate patch locations on these paths. For each candidate patch location, it generates and inserts a patch, and tests the patched program. The test result indicates the extent of side-effect caused by each patch. We evaluate the prototype of \Sysname{} on real world vulnerabilities and find that the mitigation patches generated by it can effectively mitigate these vulnerabilities with minimum side-effects.



\bibliographystyle{splncs04}
\bibliography{bibfile}

\end{document}